\documentclass{PoS}

\usepackage[utf8]{inputenc}
\usepackage{amsmath}
\usepackage{bm,braket}

\usepackage{caption}
\usepackage{subcaption}

\title{Transverse Parton Distribution Functions at Next-To-Next-To-Leading-Order}

\ShortTitle{TPDFs at NNLO}

\author{Thomas Gehrmann\\
  Institute for Theoretical Physics, University of Zürich, 8057 Zürich, Switzerland\\
  E-mail: \email{thomas.gehrmann@physik.uzh.ch}}

\author{\speaker{Thomas Lübbert}\footnote{Current address: II.\ Institute for Theoretical Physics, University of Hamburg, 22761 Hamburg, Germany.}\\
  Institute for Theoretical Physics, University of Zürich, 8057 Zürich, Switzerland\\
  E-mail: \email{tluebber@physik.uzh.ch}}

\author{Li Lin Yang\\
  School of Physics and State Key Laboratory of Nuclear Physics and Technology, Peking University, Beijing 100871, China\\
  E-mail: \email{yanglilin@pku.edu.cn}}

\abstract{We present a perturbative calculation of the transverse parton distribution functions in all partonic channels up to next-to-next-to-leading order based on a gauge invariant operator definition. We demonstrate for the first time that such a definition works beyond the first non-trivial order. We extract the coefficient functions relevant for a next-to-next-to-next-to-leading logarithmic $q_T$ resummation in a large class of processes at hadron colliders.}

\FullConference{11th International Symposium on Radiative Corrections (Applications of Quantum Field Theory to Phenomenology) (RADCOR 2013),\\
		22-27 September 2013\\
		Lumley Castle Hotel, Durham, UK }

\newcommand{\nb}{\bar{n}}

\newcommand{\Acal}{\mathcal{A}}
\newcommand{\Bcal}{\mathcal{B}}

\newcommand{\lp}{L_\perp}
\newcommand{\xp}{x_\perp}

\newcommand{\tcdot}{\!\cdot\!}
\newcommand{\as}{\alpha_s}

\allowdisplaybreaks[1]

\begin{document}

\section{Introduction}

For many processes at hadron colliders in which heavy final states are produced the transverse momentum spectrum is phenomenologically relevant. Examples of such processes are the production of Higgs bosons, vector bosons or Drell-Yan pairs of high invariant mass. To consistently describe the differential cross sections at values of the transverse momentum much smaller than the invariant mass, one needs to resum the corresponding large logarithms to all orders in perturbation theory. Pioneering work has been performed by the authors of \cite{Collins:1984kg}. In \cite{Becher:2010tm, Becher:2012yn} a framework based on the soft-collinear effective theory (SCET) of quantum chromodynamics (QCD) has been presented, which allows the proof of all order factorization theorems and renders gauge invariant operator definitions of the objects of interest. These allow direct high order determinations of the relevant objects from first principles.

The process independent objects of our interest are the transverse parton distribution functions (TPDFs). The parton-to-parton version of these we will determine in perturbation theory to next-to-next-to-leading order (NNLO). To control light-cone singularities appearing in the calculation we use the analytic regulator as suggested in \cite{Becher:2011dz}. The TPDFs are generalizations of normal PDFs, which describe the distribution of partons with specified energy fraction and transverse separation inside the considered hadron. The additional description of the transverse scale is required, since the transverse momentum of the final state which we wish to resolve is generated from recoiling against the initial state radiation.

If this transverse scale resides in the perturbative regime, the TPDFs can be expressed as Mellin convolutions of normal PDFs with perturbative matching kernels. We extract these kernels up to NNLO. They are among others required
for a next-to-next-to-next-to-leading-logarithmic (N$^3$LL) transverse momentum resummation. As a byproduct, we confirm the process specific $\mathcal{H}^{(2)}$ coefficients of \cite{Catani:2011kr, Catani:2012qa} and reextract the $\as^2$ contributions of the DGLAP splitting kernels.

While the high order $q_T$-resummation is one important application of the transverse factorization and the TPDFs,
they are also relevant for many other aspects in QCD. For example, they are important to describe spin or azimuthal related observables as well as to understand the spin structure of the proton and other hadrons.

\section{Framework}

Before generalizing our discussion to other processes, let us focus on Higgs production at hadron colliders through gluon fusion. As discussed in \cite{Becher:2012yn} with SCET based arguments, the corresponding differential cross section factorizes at small transverse momentum $q_T\ll m_H$ as
\begin{align}
  \label{eq:dsigma_sBsB_H}
  \frac{d^2\sigma}{dq_T^2 dy} &= \sigma_0(\mu) \, C_t^2(m_t^2,\mu) \, \big|C_S(-q^2,\mu)\big|^2 \, g_{\mu\rho}^\perp g_{\nu\sigma}^\perp \int \frac{d^2x_\perp}{2\pi} \, e^{-iq_\perp\cdot x_\perp} {\cal{S}}(x_\perp,\mu) \nonumber
  \\
  &\hspace{1.5cm} \times \Bcal^{\mu\nu}_{g/N_1}(z_1,x_\perp,\mu) \, \bar{\Bcal}^{\rho\sigma}_{g/N_2}(z_2,x_\perp,\mu) + \mathcal{O}\bigg(\frac{q_T^2}{m_H^2}\bigg) \, ,
\end{align}
with the Higgs rapidity $y$ and invariant mass $q^2 = m_H^2$ as well as the momentum fractions $z_1,z_2 = e^{\pm y}\sqrt{(m_H^2 + q_T^2)/s}$ along the direction of the hadrons $N_i$. $\sigma_0(\mu)$ is the Born level cross section. The Wilson coefficients $C_t$ and $C_S$ arise from the introduction of the effective $ggH$ operator and the matching to the SCET operator. They are given in \cite{Becher:2006mr, Ahrens:2008nc}. The soft function $\cal{S}$ is a correlator of soft Wilson lines. The objects of our main interest are the TPDFs $\Bcal$ and $\bar{\Bcal}$ of the collinear and anti-collinear region, respectively. Their naive form appearing above can be represented by the operator matrix element \cite{Becher:2012yn}
\begin{align}
  \label{eq:Bcal}
  \Bcal_{g/N}^{\mu\nu}(z,x_\perp,\mu) &= -\frac{z \,\nb \tcdot p}{2\pi} \int dt \, e^{-izt \nb \cdot p} \, \sum_X 
  \braket{N(p) | \Acal_{\perp}^\mu(t\nb+x_\perp) | X} \braket{X | \Acal_\perp^\nu(0) | N(p)} \, ,
\end{align}
where the sum is over all intermediate states $X$. $\Acal^\mu$ are effective gluon fields as defined in \cite{Hill:2002vw} which are dressed by Wilson lines to guarantee local gauge invariance. The transverse vectors are orthogonal to the light-cone vectors $n$ and $\nb$ which are specified by the momenta $p^\mu= (\nb \cdot p/2) n^\mu$ and $\bar{p}^\mu= (n \cdot \bar{p}/2) \nb^\mu$ of the colliding hadrons and fulfill $n \cdot \nb = 2$. To describe the partons coming from the opposite direction, we need the function $\bar{\Bcal}_{g/N}^{\mu\nu}$ which is defined by substituting $p,n \leftrightarrow \bar{p},\nb$ in the formula above.

In order to regularize the light-cone singularities when evaluating higher order corrections for these matrix elements, we introduce a factor $(\nu / n \cdot k)^\alpha$ for each emitted parton with momentum $k$ following \cite{Becher:2011dz}. Here $\nu$ is an unphysical scale associated with the regulator. 
This prescription combined with dimensional regularization regulates all singularities. In this scheme the soft function $\mathcal{S}$ in Eqn.~\eqref{eq:dsigma_sBsB_H} reduces to a trivial factor of unity.
While in Eqn.~(\ref{eq:Bcal}) the function $\Bcal$ formally only depends on a single physical scale $x_T^2=-\xp^2$, the additional regularization introduces an anomalous dependence on the scale of the hard scattering, $q^2$. Field theoretically, this is due to the breaking of the rescaling invariance of the SCET Lagrangian by the additional regulator, and was called ``collinear anomaly'' in \cite{Becher:2010tm}. The dependence on $q$ can be obtained by studying the dependence on the unphysical scale $\nu$, which leads to a refactorization of the form \cite{Becher:2010tm, Becher:2012yn}
\begin{align}
  \label{eq:refac}
  \Bcal_{g/N_1}^{\mu\nu}(z_1,x_\perp,\mu) \, \bar{\Bcal}_{g/N_2}^{\rho\sigma}(z_2,x_\perp,\mu) &= \left( \frac{x_T^2q^2}{4e^{-2\gamma_e}} \right)^{-F^{\text{b}}_{gg}(x_\perp,\mu)} \, B_{g/N_1}^{\text{b},\mu\nu}(z_1,x_\perp,\mu) \, B_{g/N_2}^{\text{b},\rho\sigma}(z_2,x_\perp,\mu) \, ,
\end{align}
where on the left hand side poles in $\alpha$ which can be present in both individual factors cancel between them and we then set $\alpha$ to 0. The right hand side is then free of both $\alpha$ and the scale $\nu$. We identify the anomaly coefficient $F$ and the two proper TPDFs $B$, which do not depend on the hard scale $q$.
The superscript ``b'' indicates that these are bare quantities which we regulate in dimensional regularization with $d=4-2\epsilon$ dimensions. The ultra-violet (UV) poles are removed by operator renormalization which takes the form
\begin{align}
  \label{eq:renorm_B}
  B_{g/N}^{\text{b},\mu\nu}(z,x_\perp) &= Z^B_g(x_\perp,\mu) \, B_{g/N}^{\mu\nu}(z,x_\perp,\mu) \, ,
  \\
  \label{eq:renorm_F}
  F^{\text{b}}_{gg}(\xp) &= F_{gg}(\xp,\mu) + Z^F_g(\mu) \, .
\end{align}
We work in the $\overline{\text{MS}}$-scheme, where at each order beyond the LO, the renormalization factors $Z$ are pure poles in $\epsilon$ multiplied by the prefactor $(4\pi)^{n\epsilon}e^{-n\epsilon\gamma_E}$, with $n$ numbering the perturbative order ($n=1$ for NLO and $n=2$ for NNLO). The TPDFs are genuinely non-perturbative objects. However, for $x_T \ll 1/\Lambda_{\text{QCD}}$, they can be matched onto the usual collinear PDFs as
\begin{align}
  \label{eq:match}
  B_{g/N}^{\mu\nu}(z,x_\perp,\mu) &= \sum_{i=q,\bar{q},g} I_{g/i}^{\mu\nu}(z,x_\perp,\mu) \otimes \phi_{i/N}(z,\mu) + \mathcal{O}(x_T^2 \Lambda^2_{\text{QCD}}) \, ,
\end{align}
with perturbatively calculable coefficient functions $I$ and the convolution defined by
\begin{align}
  f(z,\cdots) \otimes g(z,\cdots) \equiv \int_z^1 \frac{d\xi}{\xi} \, f(\xi,\cdots) \, g(z/\xi,\cdots) \, .
\end{align}
The renormalization of the collinear PDFs has the form
\begin{align}
  \label{eq:renorm_phi}
  \phi_{i/N}^{\text{b}}(z) &= \sum_{j=q,\bar{q},g} Z^\phi_{i/j}(z,\mu) \otimes \phi_{j/N}(z,\mu) \, .
\end{align}
Together with the renormalization of the TPDFs it implies the renormalization 
\begin{align}
  I^{b,\mu\nu}_{g/j}(z,\xp,\mu) &= Z^{B}_g(\xp,\mu)\sum_k I^{\mu\nu}_{g/k}(z,\xp,\mu)\otimes \phi_{k/j}(z,\mu) \, .
  \label{eq:renorm_I}
\end{align}
of the splitting kernels. The identification of the parton-to-parton PDFs in the last equation already uses one of our results which is that in our approach the perturbative corrections to the bare parton-to-parton PDFs vanish and therefore the renormalized parton-to-parton PDFs are the Mellin inverses of their renormalization kernels.

The splitting kernels obey the renormalization group (RG) equation
\begin{align}
  \label{eq:RGE_I}
  \frac{d}{d\ln \mu}I_{g/j}^{\mu\nu}(z,\xp,\mu) = \Big[ \Gamma^g_{\text{cusp}}(\alpha_s) L_\perp - 2\gamma^g(\alpha_s) \Big] I_{g/j}^{\mu\nu}(z) - 2 \sum_k  I_{g/k}^{\mu\nu}(z) \otimes P_{k/j}(z) \, ,
\end{align}
which is implied by the RG invariance of the cross section and the RG equations of the Wilson coefficients and the PDFs. The latter introduce the well known DGLAP splitting kernels $P_{k/j}$. Above, $\Gamma^g_{\mathrm{cusp}}$ is the cusp anomalous dimension in the adjoint representation, $\gamma^g$ the gluon anomalous dimension,  $L_\perp=\ln(x_T^2\mu^2e^{2\gamma_e}/4)$ and we suppressed the scale dependences on the right hand side.

The rank-2 tensor $I^{\mu\nu}$ can be decomposed into its two independent components via
\begin{align}
  I^{\mu\nu}_{g/i}(z,x_\perp,\mu) = \frac{g^{\mu\nu}_\perp}{2} \, I_{g/i}(z,L_\perp,\as) + \left( \frac{g^{\mu\nu}_\perp}{2} + \frac{x_\perp^\mu x_\perp^\nu}{x_T^2} \right) I'_{g/i}(z,L_\perp,\as) \, .
\end{align}
The contribution of the second Lorentz structure to transverse momentum resummation in gluon fusion processes was first pointed out in \cite{Catani:2010pd}. A similar decomposition holds for the bare functions, such as $\Bcal^{\mu\nu}_{g/i}$, for which we have
\begin{align}
  \Bcal^{\mu\nu}_{g/i}(z,x_\perp,\mu) = \frac{g^{\mu\nu}_\perp}{d-2} \, \Bcal_{g/i}(z,L_\perp,\as) + \left( \frac{g^{\mu\nu}_\perp}{d-2} + \frac{x_\perp^\mu x_\perp^\nu}{x_T^2} \right) \Bcal'_{g/i}(z,L_\perp,\as) \, .
\end{align}
The two functions can be projected out using
\begin{align}
  \Bcal_{g/i}(z,L_\perp,\as) &= g_{\perp\mu\nu} \,  \Bcal^{\mu\nu}_{g/i}(z,x_\perp,\mu) \, , \nonumber
  \\
  \Bcal'_{g/i}(z,L_\perp,\as) &= \frac{1}{d-3} \left( g_{\perp\mu\nu} + (d-2) \, \frac{x_{\perp\mu} x_{\perp\nu}}{x_T^2} \right) \Bcal^{\mu\nu}_{g/i}(z,x_\perp,\mu) \, .
\end{align}
Applying these projectors to the equations of this section, one straightforwardly finds the corresponding equations for the individual tensor structures.

Factorization theorems as Eqn.~\eqref{eq:dsigma_sBsB_H} hold for the production of any color neutral final state with high invariant mass at small transverse momentum. In general the Wilson coefficients change between different processes. For $gg$ initiated processes, also the tensor structure of the hard function can change. For processes initiated by $q\bar{q}$ annihilation, the hard function and the corresponding TPDFs are scalars. In their naive form, the quark TPDF can be represented by the matrix element
\begin{align}
  \label{eq:Bcal_q}
  \Bcal_{q/N}(z,\lp,\as) &= \frac{1}{2\pi} \int dt \, e^{-izt \nb \cdot p} \, \sum_X \frac{\not\!\nb_{\alpha\beta}}{2} \, \braket{N(p) | \bar{\chi}_\alpha(t\nb+x_\perp) | X} \braket{X | \chi_\beta(0) | N(p)} \, ,
\end{align}
where $\chi$ is an effective quark field dressed with collinear Wilson lines to guarantee local gauge invariance. The TPDFs for anti-quarks are given by the same expression, with the argument $t\nb+x_\perp$ carried by the field $\chi_\beta$. All points we discussed for the gluon case also hold for the (anti)-quark case with the appropriate adjustments which are essentially the removal of the tensor indices and the replacement of gluons by (anti)-quarks. Much of the relevant equations have already been provided in \cite{Gehrmann:2012ze}.

\section{Perturbative Calculation and Results}

The main subject of this talk is to calculate the NNLO corrections of the coefficient functions $I$ in Eqn.~\eqref{eq:match}, i.e.\ the coefficients $I^{(2)}$ in the expansion
\begin{align}
  I_{i/j}(z,\lp,\as) = \sum_{n=0}^\infty \left( \frac{\as}{4\pi} \right)^n I^{(n)}_{i/j}(z,\lp) \, .
\end{align}
The first step is to evaluate the matrix elements in Eqns.~(\ref{eq:Bcal}, \ref{eq:Bcal_q}) with the hadronic state $N$ replaced by some partonic state. With a gauge-invariant definition, we are free to choose any gauge to perform the calculation. We have checked that the results are the same in both the Feynman gauge and the light-cone gauge. The calculation in the light-cone gauge is particularly straightforward since the (anti)-collinear Wilson lines reduce to trivial factors of unity and the gauge-invariant effective fields become the normal (anti)-quark and gluon fields. From the results of the matrix elements, the $I$ functions as well as the $F$ functions can be extracted through the refactorization in Eq.~(\ref{eq:refac}) and the matching in Eq.~(\ref{eq:match}). 

In the calculation, we need to deal with three kinds of singularities. The light-cone singularities are regularized by the analytic regulator, and are absent in the final TPDFs. The other two, namely the ultraviolet (UV) and infrared (IR) singularities, are both regularized by dimensional regularization, and manifest themselves as poles in $\epsilon$. While $B$ contains IR-poles, for $I$ they are removed in the matching step \eqref{eq:match}. This step also allows the extraction of the renormalized PDFs up to endpoint contributions in terms of the IR poles. From those the DGLAP splitting kernels can be obtained. The endpoint contributions which appear in the parton-diagonal contributions are fixed by physical constraints on the integrals containing those kernels over the momentum fraction $z$.

\begin{figure}[tp]
\centering
  \begin{subfigure}[b]{0.16\textwidth}
   \includegraphics[width=\textwidth]{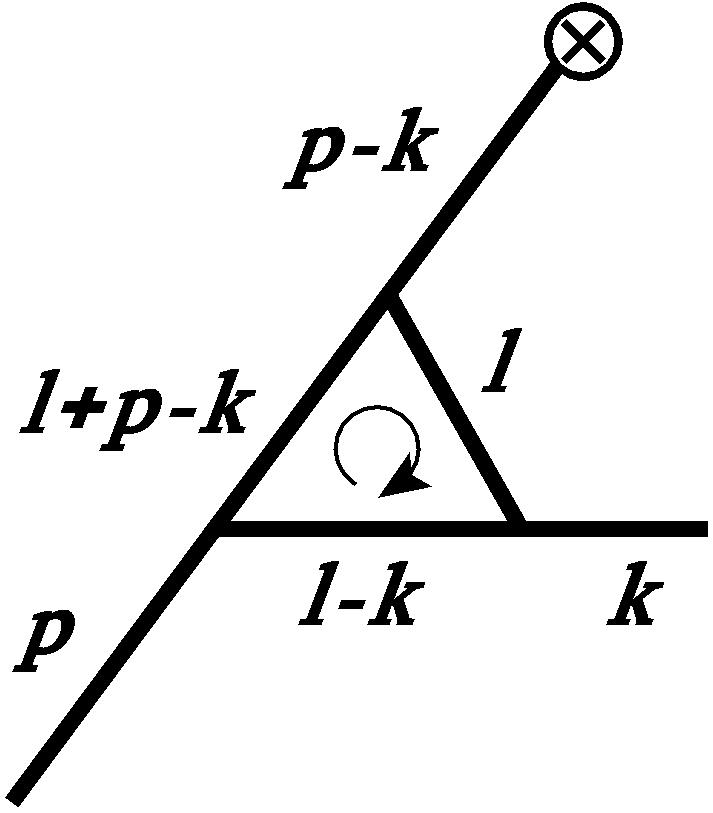}
   \caption{ }
  \end{subfigure}
\hspace{0.2cm}
  \begin{subfigure}[b]{0.135\textwidth}
   \includegraphics[width=\textwidth]{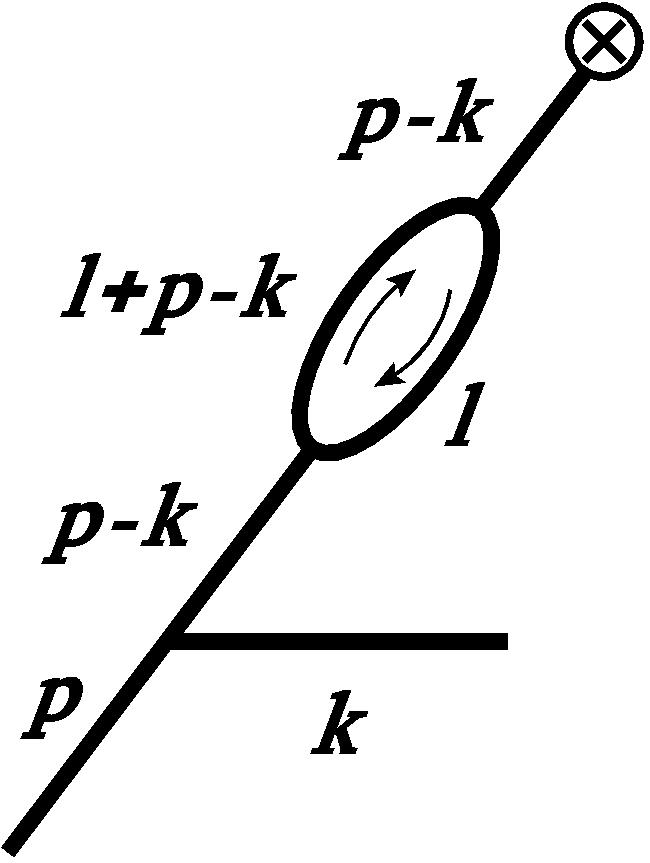}
   \caption{ }
  \end{subfigure}
\hspace{0.4cm}
  \begin{subfigure}[b]{0.15\textwidth}
   \includegraphics[width=\textwidth]{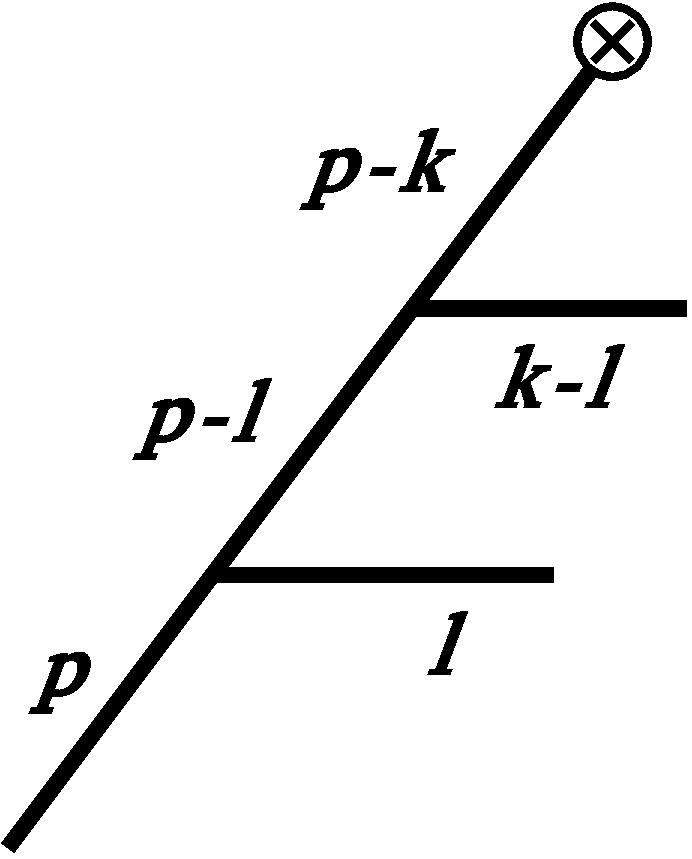}
   \caption{ }
   \end{subfigure}
\hspace{0.2cm}
  \begin{subfigure}[b]{0.15\textwidth}
   \includegraphics[width=\textwidth]{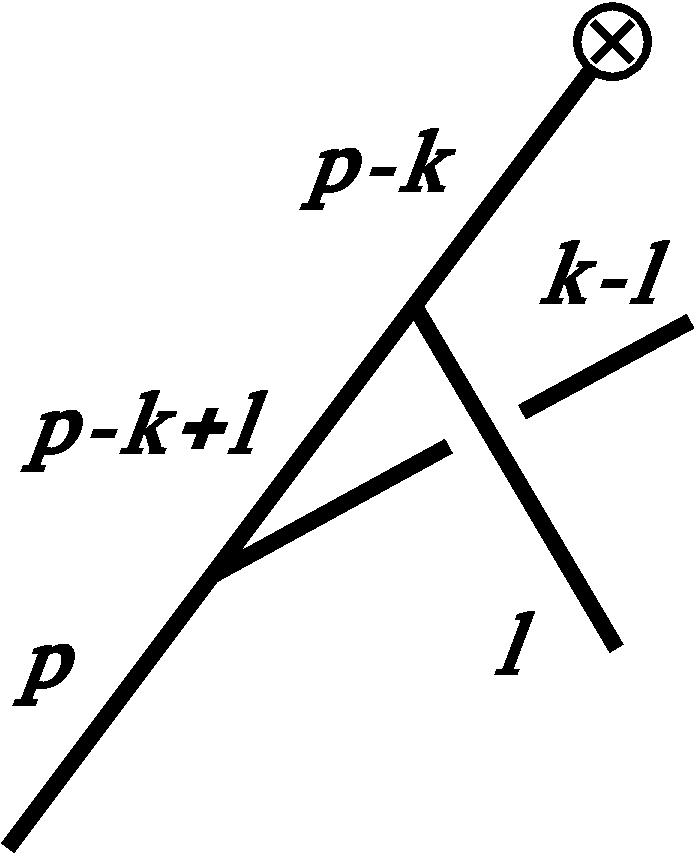}
   \caption{ }
  \end{subfigure}
\hspace{0.2cm}
  \begin{subfigure}[b]{0.16\textwidth}
   \includegraphics[width=\textwidth]{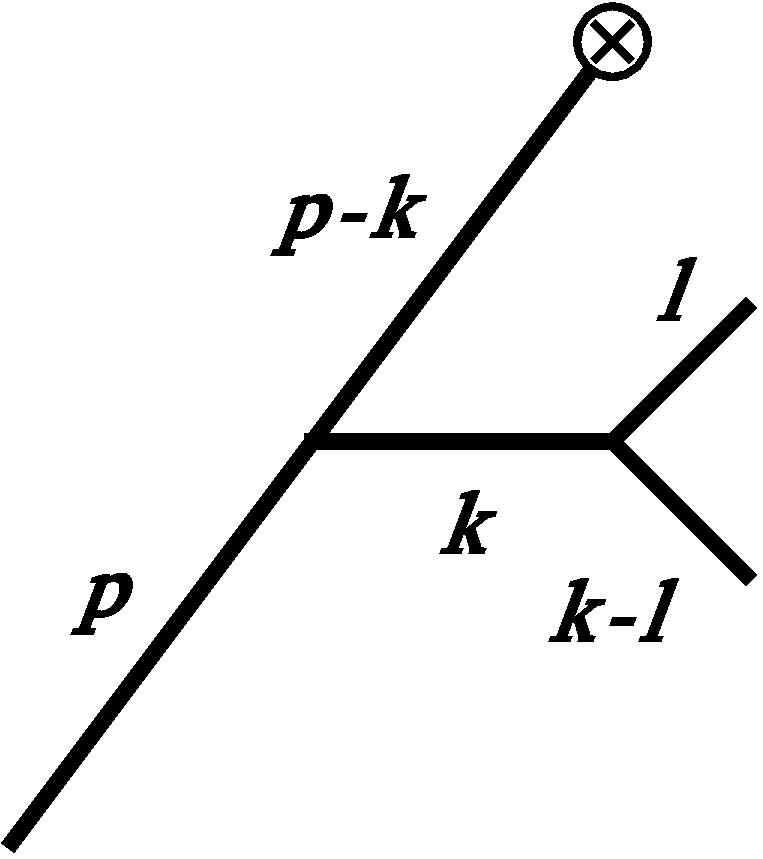}
   \caption{ }
  \end{subfigure}
  \caption{Amplitude topologies of the two NNLO contributions. The virtual-real (a,b) and double real (c,d,e) contribution.}
 \label{fig:topologies}
\end{figure}

The LO expression is trivial and just corresponds to the absorption of the incoming parton by the effective quark or gluon field represented by the vertex $\bm{\otimes}$. The NLO expression contains contributions from the square of the diagram obtained from Fig.~\ref{fig:topologies}(b) after removing the loop and using appropriate choices of partons. The unresolved parton carries momentum $k$, the incoming parton momentum $p$. The integrals appearing here can be solved in closed form as functions of the two regulators $\epsilon$ and $\alpha$. Our expanded results on the splitting kernels and anomaly coefficients confirm the results of \cite{Becher:2010tm, Becher:2012yn}. Purely virtual corrections do not contribute as they lead to scaleless integrals vanishing in dimensional regularization.

At NNLO, the TPDFs receive contributions from double real and virtual-real diagrams, which are given in Fig.~\ref{fig:topologies}. Two different 1-loop amplitude topologies with unspecified partons are depicted in Fig.~\ref{fig:topologies}(a,b). By shrinking lines to single points, the remaining amplitude subtopologies are obtained. For the virtual-real contribution, these diagrams are combined with the NLO diagram described above. The VR diagrams contain divergences requiring coupling constant renormalization, for which we include diagrams with the loop replaced by a counter-term.

The most difficult contribution is the double real one. The relevant diagrams without specified partons are given in Fig.~\ref{fig:topologies}(c,d,e) where $l$ and $k-l$ are the momenta of the unresolved partons. By shrinking propagators, the remaining amplitude topology can be identified. For the double real NNLO contribution, these diagrams are combined with each other.

From the calculation and matching steps outlined above, we can extract the final results. The NNLO anomaly coefficients are
{\small
\begin{align}
  \frac{F^{(2)}_{q\bar{q}}(x_\perp,\mu)}{C_F} = \frac{F^{(2)}_{gg}(x_\perp,\mu)}{C_A} = C_A \Big[ \frac{808}{27} -28\zeta_3 +\frac{268}{9}L_\perp - 8\zeta_2 L_\perp + \frac{22}{3} L_\perp^2 \Big] - T_F N_f \Big[ \frac{224}{27} + \frac{80}{9}L_\perp + \frac{8}{3} L_\perp^2 \Big] \, . \nonumber
\end{align}
}
The results of the matching kernels will be presented in terms of harmonic polylogarithms $H_{\vec{a}} \equiv H_{\vec{a}}(z)$ introduced in \cite{Remiddi:1999ew}, $\zeta$-values as well as the functions
\begin{gather*}
  \tilde{p}_{qq}(z) = 2 \, \frac{1+z^2}{(1-z)_+} \, , \quad \tilde{p}_{qg}(z) = 2 \left[ z^2 + (1-z)^2 \right] ,
  \\
  \tilde{p}_{gg}(z) = 4 \left[ \frac{z}{(1-z)_+} + \frac{1-z}{z} + z(1-z) \right] , \quad \tilde{p}_{gq}(z) = 2 \, \frac{1+(1-z)^2}{z} \, ,
\end{gather*}
which are related to the lowest order DGLAP splitting kernels by removing the color factors and the $\delta$-function terms. To reduce the results to a compact size,
we will only give their scale-independent parts, i.e.\ their results at $L_\perp = 0$ which are obtained for $\mu=\mu_x\equiv 4e^{-2\gamma_E}/x_T^2$. The corresponding expressions at $\mu \neq \mu_x$, containing powers of $L_\perp$, can be straightforwardly obtained from these expressions by solving the RG equations of $I_{i/j}$.

The scale independent part of the NNLO gluon-to-gluon kernel is given by
{\small 
\begin{align}
  \nonumber
  &I_{g/g}^{(2)}(z,0) = C_A^2 \bigg\{ \delta(1-z) \bigg[ \frac{25}{4} \zeta_{4} - \frac{77}{9} \zeta_{3} - \frac{67}{6} \zeta_{2} + \frac{1214}{81} \bigg]
+ \tilde{p}_{gg}(z)
  \bigg[ -H_{0,0,0} + 2H_{0,1,0} + 2H_{0,1,1} - 2H_{1,0,0} 
  \\
  &+ 2 H_{1,0,1} + 2 H_{1,1,0} + 13 \zeta_{3} - \frac{202}{27} \bigg]
+ \tilde{p}_{gg}(-z)
  \bigg[ -4 H_{ -1, -1,0} + 2 H_{ -1,0,0} + 4 H_{0, -1,0} - H_{0,0,0} - 2 H_{0,1,0} \nonumber
  \\
  &- 2H_{-1} \zeta_{2} + \zeta_{3} \bigg]
 + \bigg[ -16(1+z) H_{0,0,0} + \frac{2(25-11z+44z^2)}{3} H_{0,0}  + \frac{8(1-z)(11-z+11z^2)}{3z} \big( H_{1,0} +\zeta_{2} \big) \nonumber
  \\
  &- \frac{2z}{3} H_{1} - \frac{(701+149z+536z^2)}{9} H_{0} + \frac{4(-196+174z-186z^2+211z^3)}{9z} \bigg] \bigg\} + C_AT_FN_f \bigg\{ \delta(1-z) \bigg[ \frac{28}{9} \zeta_{3} \nonumber
  \\
  &+ \frac{10}{3} \zeta_{2} - \frac{328}{81} \bigg]
  + \frac{56}{27} \tilde{p}_{gg}(z)
  + \bigg[ \frac{8(1+z)}{3} H_{0,0} + \frac{4z}{3} H_{1} +\frac{4(13+10z)}{9} H_{0} - \frac{4(-65+54z-54z^2+83z^3)}{27z} \bigg] \bigg\} \nonumber
  \\
  &+ C_FT_FN_f \bigg\{ 8(1+z) H_{0,0,0} + 4(3+z) H_{0,0} + 24(1+z) H_{0} - \frac{8(1-z)(1-23z+z^2)}{3z} \bigg\} \, .
\end{align}
}
The quark-to-gluon kernel is obtained as
{\small
\begin{align}
  \nonumber
  &I_{g/q}^{(2)}(z,0) = C_FC_A \bigg\{
\tilde{p}_{gq}(z)
  \bigg[ 2H_{1,1,1} + 2H_{0,1,1} + 2H_{1,0,1} + 2H_{1,1,0} + 4H_{0,1,0} - 2H_{1,0,0} - \frac{11}{3} H_{1,1} \nonumber
  \\
  &+ \frac{22}{3} \big( H_{1,0} + \zeta_{2} \big) + \frac{76}{9} H_{1} + 12 \zeta_{3} - \frac{790}{27} \bigg]
+ \tilde{p}_{gq}(-z)
  \bigg[ -4 H_{ -1, -1,0} + 2 H_{ -1,0,0} + 4 H_{0, -1,0} - 2 H_{ -1} \zeta_{2} \bigg] \nonumber
  \\
  &+ \bigg[ -4(2+z) H_{0,0,0} + 16 H_{0,1,0} + 4 z H_{ -1,0} + 4 z H_{0,1} + 4 z H_{1,1} - \frac{8(1+z+2z^2)}{3} H_{1,0} + \frac{2(36+9z+8z^2)}{3} H_{0,0} \nonumber
  \\
  &- \frac{22z}{3} H_{1} - \frac{2(249-6z+88z^2)}{9} H_{0} - 8 \zeta_{3} - \frac{2(4+13z+8z^2)}{3} \zeta_{2}  + \frac{4(1+127z+152z^2)}{27} \bigg] \bigg\} \nonumber
  \\
  &+ C_F^2 \bigg\{
\tilde{p}_{gq}(z)
  \bigg[ -2 H_{1,1,1} + 3 H_{1,1} - 8 H_{1} \bigg]
  + \bigg[ 2(2-z) H_{0,0,0} - (4+3z) H_{0,0} - 4 z H_{1,1} + 6 z H_{1} - 5(3-z) H_{0} \nonumber
  \\
  &+ (10-z) \bigg] \bigg\} + C_FT_FN_f \bigg\{
\tilde{p}_{gq}(z)
  \bigg[ \frac{4}{3} H_{1,1} - \frac{20}{9} H_{1} + \frac{112}{27}\bigg]
  + \bigg[ \frac{8z}{3} H_{1} - \frac{40z}{9} \bigg] \bigg\} \, ,
\end{align}
}
while the gluon-to-quark kernel reads
{\small
\begin{align}
  \nonumber
  &I_{q/g}^{(2)}(z,0) = C_AT_F \bigg\{
\tilde{p}_{qg}(z)
  \bigg[ 2 H_{1,0,1}  + 2 H_{1,1,0} - 2 H_{1,1,1} + 2 H_{1,1} - \frac{22}{3} H_{0,0} + \frac{22}{3} \big( H_{1,0} + \zeta_{2} \big) + \frac{68}{9} H_{0} + 2 H_{1} \nonumber
  \\
  &- \frac{149}{27} \bigg]
+ \tilde{p}_{qg}(-z)
  \bigg[ -4 H_{-1,-1,0} + 2 H_{ -1,0,0} + 4 H_{0, -1,0} + 2 H_{ -1,0} - 2 H_{ -1} \zeta_{2} \bigg]
  + \bigg[ 4(1+2z) H_{0,0,0} - 16 z H_{0,1,0} \nonumber
  \\
  &+ \frac{2(19-32z)}{3} H_{0,0} - 4 H_{ -1,0} - 4 H_{1,1} - \frac{4(13-38z)}{9} H_{0} - \frac{4(4+5z+2z^2)}{3z} \big( H_{1,0} + \zeta_{2} \big) + 2(-2+z) H_{1} + 8 z \zeta_{3} \nonumber
  \\
  &+ 8z \zeta_{2}  + \frac{2(172-166z+89z^2)}{27z} \bigg] \bigg\} + C_FT_F \bigg\{
\tilde{p}_{qg}(z)
  \bigg[ 2 H_{1,1,1} - 2 H_{1,0,0} + 2 H_{0,1,1} - 2 H_{0,0,0} - 2 H_{1,1} - 2 H_{1,0} \nonumber
  \\
  &- 2 H_{0,1} - 2 H_{0,0} - 2 H_{1} - 2 H_{0} + 14 \zeta_{3} + 3 \zeta_{2} - 18 \bigg]
  + \bigg[ 2(1-2z) H_{0,0,0} + (5+4z) H_{0,0} + 4 H_{0,1} + 4 H_{1,0} \nonumber
  \\
  &+ 4 H_{1,1} + 2(2-z) H_{1} + (12+7z) H_{0} - 6 \zeta_{2} + (23+3z) \bigg] \bigg\} \, .
\end{align}
}
The matching kernel $I_{q/q}^{(2)}$ for a quark evolving to a quark of the same flavor was already calculated by us in \cite{Gehrmann:2012ze}. For a quark evolving to a quark (or anti-quark) of different flavor, it is instead given by
{\small
\begin{align}
  &I_{q'/q}^{(2)}(z,0) = C_FT_F \bigg\{ 4 (1+z) H_{0,0,0} - \frac{2(3+3z+8z^2)}{3} H_{0,0} - \frac{8(1-z)(2-z+2z^2)}{3z} \big( H_{1,0} + \zeta_{2} \big) \nonumber
  \\
  &+ \frac{4(21-30z+32z^2)}{9} H_{0} + \frac{2(1-z)(172-143z+136z^2)}{27z} \bigg\} \, .
\end{align}
}
For a quark evolving to an anti-quark of the same flavor, the matching kernel is
{\small
\begin{align}
  \nonumber
  &I_{\bar{q}/q}^{(2)}(z,0) = \big(C_FC_A-2C_F^2\big) \bigg\{
\tilde{p}_{qq}(-z)
  \bigg[ 4 H_{ -1, -1,0} - 2 H_{ -1,0,0} - 4 H_{0,-1,0} + 2 H_{0,1,0} + H_{0,0,0} + 2 H_{ -1} \zeta_{2} - \zeta_{3} \bigg] \nonumber
  \\
  &+ \bigg[ 4(1-z) H_{1,0} + 4(1+z) H_{-1,0} - (3+11z) H_{0} + 2(3-z) \zeta_{2} - 15(1-z) \bigg] \bigg\} +I_{q'/q}^{(2)}(z,0) \, .
\end{align}
}
All other splitting kernels $I_{i/j}^{(2)}$ are related by flavor or charge conjugation symmetry to the results provided here and in \cite{Gehrmann:2012ze}. Most of the symmetries are already respected by not explicitly introducing $N_f$ different flavors, but only providing a quark $q$ of unspecified (but same) flavor and a quark $q'$ of different flavor. Another set of relations is obtained from the equality $I_{\bar{\imath}/\bar{\jmath}} = I_{i/j}$. Moreover, up to NNLO one has $I_{\bar{q}'/q}=I_{q'/q}$. As a check, we extended our calculation to such additional combinations of partons and found agreement.

The process-independent matching kernels can among others be applied to N$^3$LL $q_T$ resummation in a wide range of processes at hadron colliders, in which a color-neutral final state with high invariant mass and small transverse momentum is produced. Examples of such processes are the production of a Drell-Yan pair, individual or multiple vector bosons or a Higgs boson. For the last case, also the NLO $I'_{g/j}$ functions are needed which we also determined. Since the LO of these function vanishes, the NNLO $I'_{g/j}$ functions are needed for N$^3$LL accuracy only for gluon-gluon initiated processes with off-diagonal tensor structure. They can be calculated in a similar way as the kernels presented here.

We applied several non-trivial checks to our results. The first point we observe is that the results are consistently free of poles in the two regulators $\alpha$ and $\epsilon$. Moreover, they only depend on mass scales in terms of $L_\perp$. Especially, as required by consistency, they depend neither on the scale $\nu$ associated to the analytic regulator nor on the hard scale $q^2$. These points are not only a strong confirmation of our results, but also of the consistency of the whole framework. Our calculation also demonstrates that the analytic regulator of \cite{Becher:2011dz} can be practically and consistently applied in involved NNLO calculations. Also the renormalization in terms of Eqn.~\eqref{eq:renorm_I} was highly constraining, as for the whole set of splitting kernels all $\epsilon$ poles could be removed by only specifying the two multiplicative factors $Z^B_i$, while the expressions for the parton-to-parton PDFs had been already implied by the DGLAP splitting kernels.
Furthermore, we confirmed that the functions $I_{i/j}(z,\lp,\as)$ obey the RG equation \eqref{eq:RGE_I}, which is yet another very strong test of our results.

In addition to these tests, we compared our results to literature. We could directly compare our NLO results for the matching kernels as well as the NNLO anomaly coefficients to \cite{Becher:2010tm, Becher:2012yn} and found agreement.
Using our results of the matching kernels and the Wilson coefficients appearing in the factorization theorems for Drell-Yan and Higgs production, respectively, we can confirm the corresponding $\mathcal{H}$ coefficients of \cite{Catani:2011kr, Catani:2012qa} up to $\as^2$ by using the relations
\begin{align}
  \mathcal{H}^{DY}_{q\bar{q}\leftarrow jk}(z,\as) &= \big|C_V(-M^2-i\epsilon,M)\big|^2 I_{q/j}(z,0,\as) \otimes I_{\bar{q}/k}(z,0,\as) \, ,
  \\
  \mathcal{H}^{H}_{gg\leftarrow jk}(z,\as,L_h) &= C_t^2(m_t^2,m_h) \big|C_S(-m_h^2-i\epsilon,m_h)\big|^2 \nonumber
  \\
  &\times \big[ I_{g/j}(z,0,\as) \otimes I_{g/k}(z,0,\as) + I_{g/j}'(z,0,\as) \otimes I_{g/k}'(z,0,\as) \big] \, .
\end{align}
Since the determination in \cite{Catani:2011kr, Catani:2012qa} is done in a completely different framework the agreement is yet another very strong check for our as well as their results and frameworks.

\medskip

In conclusion, we have calculated the perturbative parton-to-parton TPDFs at NNLO based on a gauge invariant operator definition with an analytic regulator. We demonstrate for the first time that such a definition works beyond the first non-trivial order. We extract from our calculation the coefficient functions relevant for a N$^3$LL $q_T$ resummation. 
Our results can be applied to all processes yielding a colorless final state, provided the NNLO virtual corrections are known. Combined with the work \cite{Zhu:2012ts}, our results could also be applied for $t\bar{t}$ production. For gluon-gluon initiated processes with a general spin structure, in addition to the results presented here the NNLO results for the second tensor structure of the gluon TPDFs are required for a N$^3$LL $q_T$ resummation. The corresponding results will be presented in a forthcoming article.

\medskip

\textit{Acknowledgments}: This work was supported in part by the Schweizer Nationalfonds under grant 200020-141360/1, by the Research Executive Agency (REA) of the European Union under the Grant Agreement number PITN-GA-2010-264564 (LHCPhenoNet), and by the National Natural Science Foundation of China under Grant No. 11345001.

\end{document}